\documentclass[]{interact}

\usepackage{epstopdf}
\usepackage[caption=false]{subfig}

\usepackage[natbibapa,nodoi]{apacite}
\theoremstyle{plain}

\theoremstyle{definition}

\theoremstyle{remark}

\begin{document}

\articletype{Research Article}

\title{Probabilistic Multilabel Graphical Modelling of Motif Transformations in Symbolic Music}
%
\author{
\name{Ron Taieb\textsuperscript{a}\thanks{CONTACT Ron Taieb. Email: \texttt{ron.taieb@mail.huji.ac.il}}
\and Yoel Greenberg\textsuperscript{b}
\and Barak Sober\textsuperscript{a}}
\affil{\textsuperscript{a}Department of Statistics and Data Science, The Hebrew University of Jerusalem, Jerusalem, Israel;
\textsuperscript{b}Department of Musicology, The Hebrew University of Jerusalem, Jerusalem, Israel}
}

\maketitle

\begin{abstract}
Motifs often recur in musical works in altered forms, preserving aspects of their identity while undergoing local variation. This paper investigates how such motivic transformations occur within their musical context in symbolic music. To support this analysis, we develop a probabilistic framework for modeling motivic transformations and apply it to Beethoven’s piano sonatas by integrating multiple datasets that provide melodic, rhythmic, harmonic, and motivic information within a unified analytical representation.

Motif transformations are represented as multilabel variables by comparing each motif instance to a designated reference occurrence within its local context, ensuring consistent labeling across transformation families. We introduce a multilabel Conditional Random Field (CRF) to model how motif-level musical features influence the occurrence of transformations and how different transformation families tend to co-occur.

Our goal is to provide an interpretable, distributional analysis of motivic transformation patterns, enabling the study of their structural relationships and stylistic variation. By linking computational modeling with music-theoretical interpretation, the proposed framework supports quantitative investigation of musical structure and complexity in symbolic corpora and may facilitate the analysis of broader compositional patterns and writing practices.
\end{abstract}


\section{Introduction}

Many musical works are shaped by the recurrence of small ideas that appear in multiple forms over the course of a composition. A motif is typically a short and recognizable configuration of pitches, rhythms, or intervals that returns in different contexts while preserving aspects of its identity. Through repetition and variation of selected musical characteristics, motifs contribute to both local continuity and large-scale organization. The ways in which these ideas reappear and change play an important role in shaping formal structure and stylistic character.

The study of motivic development has long been an important part of music-theoretical analysis and has traditionally relied on detailed examination of individual works. Such close reading provides rich interpretive insight and remains central to analytical practice. At the same time, the increasing availability of symbolic music corpora has created new opportunities to investigate recurring structural processes across larger collections. Computational methods make it possible to trace occurrences of musical ideas systematically and to compare their realizations across contexts. Much of this work has focused on representation, similarity, and detection, often with probabilistic modeling components, while there remains room for approaches that support formal statistical inference about the structural patterns underlying motivic variation.

In this study we ask how motifs are transformed within their musical context and what structural tendencies govern how different transformations appear and interact. Composers often alter motifs while preserving aspects of their identity, allowing ideas to evolve while remaining recognizable. Understanding these processes involves examining both the musical characteristics of individual motif instances and the relationships among transformations that occur within the same local musical environment.

The main contribution of this work is to provide a statistical framework for studying how motivic ideas evolve within their musical context through patterns of transformation. By combining multilabel graphical modeling with formal statistical inference, the approach enables cumulative, corpus-level analysis of how transformation types relate to musical features and to one another, while maintaining a direct connection to music-theoretical interpretation.

To operationalize this framework, we model transformation labels using a multilabel Conditional Random Field estimated for inferential purposes rather than prediction. The model relates melodic, rhythmic, and harmonic characteristics of motif instances to transformation probabilities and captures interaction patterns among transformation families. Parameter estimation enables confidence interval construction and hypothesis testing, allowing structural tendencies to be evaluated statistically.

To study these questions empirically, we examine motivic variation in Beethoven’s piano sonatas. Within each sonata, recurring musical ideas are represented by motif classes, and each appearance of such an idea in the score is treated as a motif instance. We are interested in how instances of the same motif class relate to one another within their immediate musical context. Because variation is not explicitly annotated, we derive transformation labels by comparing each instance with a designated reference occurrence within its local musical phrase, later formalized as a segment. The data exhibit several characteristic features: many motif instances preserve their core configuration, some transformation types occur only rarely, and multiple forms of variation may occur simultaneously within a single instance. This structure results in a sparse multilabel setting in which dependencies among transformation types form an important part of the phenomenon.

We examine these patterns across stylistic periods of Beethoven’s piano sonatas, with particular attention to differences between early and middle works. By comparing both the magnitude and the direction of estimated effects across periods, the analysis allows transformation patterns and their interactions to be evaluated while accounting for differences in data availability and effective sample size. This approach connects computational modeling with music-theoretical interpretation and supports statistically grounded investigation of compositional structure and stylistic development in symbolic music corpora. The analysis reveals structured differences in how transformation patterns are organized across periods, both in their associations with musical features and in their contextual interaction structure.

\section{Related Work}

A central direction in computational musicology concerns the identification and analysis of recurring patterns in symbolic music corpora. Early work on pattern discovery established algorithmic approaches for detecting repeated structures in polyphonic scores, including the SIATEC family and related methods \citep{Meredith2006,Meredith2013}, with later developments improving computational efficiency and scalability \citep{Bjoerklund2022}. Alternative frameworks have explored different representations of musical information in order to reveal regularities, including geometric and signal-based approaches as well as multiple-viewpoint representations that capture complementary musical dimensions \citep{Meredith2016,Conklin2001}. More recent work has further addressed transformation-aware pattern discovery, allowing recurring material to be identified under operations such as transposition or temporal scaling \citep{Laaksonen2022}. Together, these studies establish a strong methodological foundation for locating and characterizing recurring musical ideas within large symbolic collections.

The increasing availability of annotated corpora has enabled the study of motivic structure across larger collections of works. The BPS-Motif dataset \citep{Hsiao2023} provides expert-annotated recurring patterns across Beethoven’s piano sonatas together with a dedicated polyphonic discovery method, which makes it a useful resource for corpus-level motivic analysis. At the same time, research on annotation agreement has emphasized the interpretive nature of structural labeling and the variability that can arise even among expert analysts \citep{Tomasevic2021}. Additional studies have examined the role and evaluation of repeated patterns in corpus-based analysis, demonstrating their usefulness for stylistic characterization and large-scale comparison \citep{Boot2016,Janssen2014}. Collectively, this line of work supports a shift from single-work analysis toward the investigation of recurring structural processes across repertoires.

Graph-based representations have also been used to model structural relationships within symbolic music. Relational and spectral approaches have been applied to motif comparison and ranking \citep{Pinto2010,Simonetta2018}, while graph-based structure analysis and changepoint methods illustrate how musical organization can be examined at multiple structural levels \citep{Hernandez2026}. More recent developments have provided general graph-processing frameworks for symbolic scores and explored representation-learning approaches that capture motif-level relationships \citep{Karystinaios2022,Karystinaios2024,Wu2023}. These studies demonstrate the flexibility of graph structures for representing local musical context and structural interaction.

Alongside these developments, statistical perspectives have begun to complement structural analysis. For example, \citep{Conklin2019} used significance testing to identify stylistically meaningful absences of musical patterns, showing how statistical inference can provide insight into compositional constraints within a corpus. Building on this emerging direction, the present work combines corpus-based motif analysis and graph-based structural representation with a probabilistic framework that supports statistical inference on transformation behavior and their interrelations.

\section{Data and Preprocessing}
\label{sec:data}

\subsection{Corpus and Data Sources}

The analysis is based on the \textit{Beethoven Piano Sonatas Dataset (BPSD)} \citep{Zeitler2024}. The dataset provides symbolic MusicXML encodings of the first movements of Beethoven’s piano sonatas and includes detailed note-level information together with functional harmony annotations derived from the BPS-FH resource \citep{Chen2018}. This provides a representation that combines note-level score information derived from MusicXML with functional harmony annotations, allowing musical events to be analyzed in relation to their tonal context.

To enable the study of motivic variation, we incorporated annotations from the \textit{BPS-Motif} dataset \citep{Hsiao2023}, which identifies recurring musical patterns and their occurrences throughout the corpus. Motivic annotations distinguish between \emph{motif classes} and their individual occurrences. Let $\mathcal{T}={1,\dots,T}$ denote the set of motif classes defined within a movement. Each class corresponds to a recurring musical pattern, while each realization of that pattern in the score is treated as a \emph{motif instance}. The resource contains more than 250 motif classes and approximately 5,000 annotated instances across the corpus.

Because BPS-Motif is distributed in a MIDI-based CSV format whereas BPSD is provided in MusicXML, a preprocessing step was performed to align motif instances with the corresponding notes in the symbolic score, which introduces representation mismatches (e.g., in pitch spelling, timing, and notation) that must be resolved during alignment. This integration produced a unified corpus in which each motif instance is linked to its full notational, harmonic, and temporal context. Full implementation details of the alignment and preprocessing procedures, together with the processed dataset and code used in the analysis, are publicly available in the accompanying GitHub repository.\footnote{\url{https://github.com/ron-taieb/beethoven-motif-transformations-crf}}

All subsequent analyses are conducted at the level of motif instances. Each instance constitutes a single observational unit and is later associated with transformation labels and aggregated musical descriptors used as covariates in the statistical modeling framework.

\subsection{Motif-Level Features}

Each motif instance was further characterized by a set of aggregated musical descriptors derived from its aligned note-level representation. These features summarize properties of the instance along multiple musical dimensions, including pitch, rhythm, harmony, and expressive markings such as dynamics and articulation.

For each instance, note-level attributes were aggregated into summary statistics such as pitch range, rhythmic variability, harmonic variation, and expressive density, defined as the average density of expressive markings (e.g., slurs, crescendi, accents) within the instance. The selected features were intended to capture general musical characteristics that may influence how a motif is varied within its local context, while avoiding redundancy. Highly correlated variables and descriptors that duplicated information already reflected in the transformation definitions were removed.

Table~\ref{tab:motif_features} summarizes the resulting feature set. While this collection reflects the musical dimensions emphasized in the present study, the framework is not restricted to a fixed set of descriptors. Alternative or additional features may be incorporated depending on the analytical goals or musical aspects of interest.

\begin{table}
\tbl{Motif-level features computed for each motif instance.}
{\begin{tabular}{p{4cm} p{9cm}} \toprule
Feature & Description \\ \midrule
Spread Harmonic Complexity & Difference between maximum and minimum harmonic complexity across chords within the instance. \\
Secondary Chord Proportion & Proportion of secondary-function chords relative to all chords. \\
Key Change Count & Number of local key changes normalized by motif duration. \\
Pitch Spread & Range (max-min) of MIDI pitch values. \\
Motif Pitch Register & Median pitch height relative to the local key center. \\
IOI Standard Deviation & Standard deviation of inter-onset intervals (rhythmic variability). \\
Silence Proportion & Fraction of silence (gaps between notes) relative to motif duration. \\
Metrical Stress Rate & Average metrical weight based on beat position within the measure. \\
Expressive Density & Mean density of expressive markings (e.g., slurs, crescendi, accents). \\
Dynamic Variability & Standard deviation of dynamic levels across the instance. \\
\bottomrule
\end{tabular}}
\label{tab:motif_features}
\end{table}

\subsection{Data Segmentation}

To evaluate motif instances within stable local musical contexts, each movement was partitioned into contiguous segments intended to approximate phrase-level structure. Segmentation was performed using a rule-based procedure that identified candidate boundaries from three types of musical cues: (i) temporal gaps in the musical texture (rests indicating true silence), (ii) sustained repetitions of the same pitch separated by at least a quarter-note duration, which often signal local closure, and (iii) harmonic events associated with cadential motion. Candidate boundaries were extracted in continuous score time and filtered using two constraints. First, boundaries falling inside annotated motif instances were removed to preserve motif integrity. Second, adjacent boundaries were required to be separated by a minimum span (eight measures) to avoid overly fragmented segments. When multiple cues occurred in close proximity, silence and repeated-pitch boundaries were preferred over harmonic candidates, as these provide stronger structural evidence and reduce spurious segmentation due to local harmonic fluctuation. The resulting segments define local contexts within which motif instances are compared, allowing variation to be assessed relative to nearby material while limiting the influence of large-scale structural drift.

\subsection{Transformation Evaluation}

\begin{figure}
\centering
\includegraphics[width=0.8\textwidth]{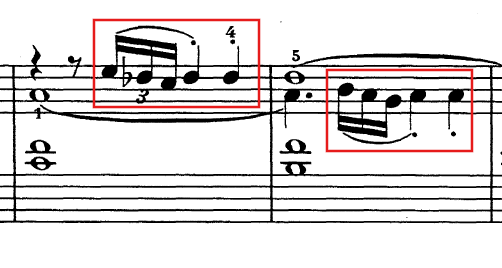}
\caption{Example of a motif instance exhibiting harmony-preserving and intervallic-preserving transformations relative to its anchor (harmonic progression $IV^{65}$ and $ii^{7}$ within a Pre-Dominant function).}
\label{fig:transformation_example}
\end{figure}

Within each segment, motif instances were evaluated relative to a local reference. For each motif class, the first occurrence in the segment was designated as the \emph{anchor}. This choice reflects the perceptual role of the initial appearance as a local prototype against which subsequent occurrences are heard as variations. When a motif class appeared only once within a segment, the most recent anchor from a preceding segment was used to preserve contextual continuity.

All other instances of the same class were compared to the anchor to determine their transformations. This choice provides a consistent reference for evaluating transformations across motif instances, although in some musical contexts variations may instead accumulate sequentially across successive occurrences rather than relating directly to the first instance. The present approach favors a stable reference point for comparison while acknowledging that both perspectives may capture meaningful aspects of motivic variation.

Because motif instances may differ in length due to ornamentation or note edits, direct one-to-one comparison is not always possible. Anchor-instance pairs were therefore aligned to establish note correspondences using pitch, metrical position (beat), and duration information, with a dynamic programming alignment \citep{bellman1966dp} applied when simple matching was insufficient. These correspondences allow note-wise comparison used to compute the transformation labels described below.

Transformation labels were designed to capture how a motif instance relates to its anchor in terms of preserved structural properties, with several definitions, particularly those concerning functional harmonic organization, drawing on established analytical concepts in tonal theory \citep{White2020}. Retrograde transformations were not evaluated because they occur only rarely in the present repertoire. Rather than describing arbitrary changes, the evaluation focuses on whether specific musical dimensions, such as melodic shape, rhythm, harmony, or interval structure, remain stable under variation. In addition to preservation-based relations, the framework also includes structural modifications such as note insertion or deletion that change the length of the instance. All labels were computed automatically from the aligned motif pairs. While the transformation families defined here reflect the musical dimensions emphasized in the present study, the modeling framework is not restricted to a fixed set of labels, and alternative transformation definitions may be incorporated depending on analytical goals. Figure~\ref{fig:transformation_example} illustrates an example of a motif instance that preserves both harmonic function and interval structure relative to its anchor.

\begin{table}
\tbl{Transformation families evaluated for each motif instance relative to its segment-level anchor.}
{\begin{tabular}{p{4cm} p{9cm}} \toprule
Transformation & Description \\ \midrule
Identity & Aligned notes share identical pitch and proportionally scaled duration, indicating the instance reproduces the anchor without structural change. \\

Contour-Preserving & The direction of successive pitch intervals (ascending/descending pattern) is maintained, while the exact interval sizes may differ. \\

Salient Leap-Preserving & Prominent melodic leaps are preserved, maintaining the locations of large pitch discontinuities within the motif. \\

Rhythm-Preserving & The temporal structure is maintained, with inter-onset intervals varying consistently under a common scaling of temporal proportions. \\

Harmony-Preserving & The sequence of harmonic functional zones (Tonic, Pre-Dominant, Dominant) remains consistent across aligned notes. \\

Intervallic-Preserving & The interval roles between successive notes (e.g., third, fifth, tritone) are preserved modulo octave, even if absolute pitch changes. \\

Note Addition/Removal & One or more notes are inserted or deleted relative to the anchor, resulting in a length difference after alignment, where only changes introducing a pitch distinct from adjacent notes are treated as structural. \\

Symmetry & The pitch structure is reorganized through inversion of melodic contour or reordering of interval content while preserving the underlying interval relationships. \\

\bottomrule
\end{tabular}}
\label{tab:transformation_families}
\end{table}

\section{Statistical Model}

\subsection{Modeling Transformation Structure}

The transformation labels described in Section~\ref{sec:data} form a multilabel outcome defined at the level of motif instances. Let \(N\) denote the number of motif instances under analysis. Let \(Y \in \{0,1\}^{N \times Q}\) denote the matrix of transformation indicators, where \(Q\) is the number of transformation families and \(Y_i^q=1\) if instance \(i\) exhibits transformation type \(q\), and let \(X \in \mathbb{R}^{N \times p}\) denote the corresponding matrix of motif-level features. The observed labels are sparse, as structurally meaningful variation is rare, and many instances show no transformation relative to their segment-level anchor.

Transformation behavior also reflects local musical context. Motif instances that occur close to one another within the same segment are not independent, but tend to exhibit coordinated patterns shaped by phrase-level compositional processes. The statistical framework must therefore account for multilabel structure, sparsity, and dependence among nearby instances.

To capture these characteristics, transformation labels are modeled using a multilabel Conditional Random Field (CRF) \citep{Lafferty2001}. The model specifies a conditional distribution of transformations given motif-level features while allowing dependencies among instances within their local context, enabling interpretable analysis of transformation tendencies and their relationships. The framework is used for descriptive and inferential analysis of transformation structure rather than for prediction.

\subsection{Graph Representation of Local Musical Context}
\label{subsec:graphical_representation}

Dependencies among motif instances are represented through an undirected graph that reflects the segment-based notion of local musical context. Within each movement, motif instances are ordered by onset time and partitioned according to the segmentation described in Section~\ref{sec:data}. Within each segment, all instances are connected, while no edges are defined across segment boundaries. The resulting structure is block-diagonal, with each segment forming an independent subgraph. This restriction is primarily motivated by computational tractability; while dependencies may in principle extend across segments (e.g., between structurally related passages), the present formulation focuses on local interactions within segments.

Each node corresponds to a motif instance and is associated with its multilabel transformation vector \(Y_i \in \{0,1\}^Q\). Edges between instances \(i\) and \(j\) within the same segment are weighted according to their ordinal proximity using a Gaussian decay function:
\begin{equation}
w_{ij} = \exp\!\left(-\frac{(i-j)^2}{\sigma^2}\right).
\label{eq:edge_weight}
\end{equation}
This weighting assigns stronger interactions to nearby occurrences while allowing the influence of more distant events to decrease smoothly. In the present analysis, the scale parameter was fixed at \(\sigma = 1\); alternative values produced qualitatively similar results. To avoid overly dense graphs, edges with negligible weights (\(<10^{-5}\), corresponding to separations of about 10 positions) were removed.

\begin{figure}
\centering
\includegraphics[width=0.9\textwidth]{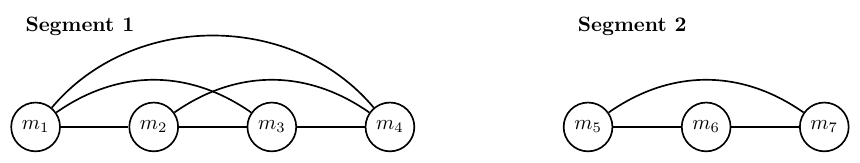}
\caption{Example of the interaction graph for two segments. Each node represents a motif instance, and edges connect all pairs of instances within the same segment. Edge strengths decay with temporal distance according to Equation~\ref{eq:edge_weight}. Each segment forms an independent subgraph, resulting in a block-diagonal structure that restricts interactions to local musical contexts.}
\label{fig:segment_graph}
\end{figure}

This construction translates phrase-level musical context into a statistical structure: motif instances that occur close together within a segment are allowed to exhibit stronger dependence, while interactions remain confined to musically coherent regions. Figure~\ref{fig:segment_graph} illustrates the resulting graph structure.

\subsection{Conditional Model Formulation}
\label{subsec:crf_formulation}

Transformation behavior is modeled using a Conditional Random Field (CRF) defined over the interaction graph described in Section~\ref{subsec:graphical_representation}. The model combines two sources of information: the influence of motif-level features on individual transformations (unary effects) and dependencies among transformations across nearby motif instances within the same segment (pairwise effects).

Let \(X\) and \(Y\) denote the feature and transformation matrices as defined in Section~\ref{sec:data}. Interactions among instances are encoded by the adjacency matrix \(\mathcal{A} \in \mathbb{R}^{N \times N}\) associated with the graph defined in Section~\ref{subsec:graphical_representation}. The matrix is block-diagonal by segment, with entries \(\mathcal{A}_{ij}\) given by symmetrically normalized Gaussian proximity weights \eqref{eq:edge_weight} and zeros for instances belonging to different segments. Consequently, no interactions are specified across segment boundaries, and the joint model factorizes over segments.

Formally, the conditional distribution is
\begin{equation}
P(Y \mid X) = \frac{1}{Z(X)}
\exp\Bigg(
\sum_{i=1}^{N} \sum_{q=1}^{Q} 
Y_i^q \, X_i^\top \alpha_{\cdot q}
\;+\;
\sum_{i,j=1}^{N} \mathcal{A}_{ij}
\sum_{q=1}^{Q} \sum_{r=1}^{Q} Y_i^q Y_j^r \, \beta_{qr}
\Bigg),
\label{eq:crf}
\end{equation}
Here \(\alpha \in \mathbb{R}^{p \times Q}\) contains feature coefficients, \(\beta \in \mathbb{R}^{Q \times Q}\) contains interaction parameters, and \(Z(X)\) is a normalization constant. The interaction matrix \(\beta\) is constrained to be symmetric (\(\beta_{qr} = \beta_{rq}\)), and the adjacency matrix \(\mathcal{A}\) is symmetric, reflecting the undirected nature of the interaction graph. Under these constraints, interactions between transformation pairs are not double-counted, ensuring that each pair contributes uniquely to the interaction term.

The two components of \eqref{eq:crf} have complementary interpretations:

\begin{itemize}
\item \textbf{Unary effects (\(\alpha\)).}  
The unary effects capture the relationship between motif-level features and transformation labels, linking musical characteristics to the likelihood of each transformation. For each transformation \(q\), the quantity \(X_i^\top \alpha_{\cdot q}\) represents a projection of the feature vector of instance \(i\) onto a learned parameter vector. This contribution is activated only when the label \(Y_i^q = 1\), so the unary term describes how local musical characteristics influence the likelihood of each transformation. If the pairwise term were omitted, the model reduces to a multilabel logistic regression in which transformation families are conditionally independent given the features.

\item \textbf{Pairwise effects (\(\beta\)).}  
The pairwise effects capture dependencies between transformation labels across nearby motif instances, reflecting how transformations tend to co-occur within local musical contexts. The second term has a quadratic form in the label variables and represents graph-weighted co-occurrence of transformations across motif instances. Each product \(Y_i^q Y_j^r\) indicates that transformation \(q\) occurs at instance \(i\) and transformation \(r\) at instance \(j\), with interactions weighted by the adjacency structure \(\mathcal{A}\). The parameter \(\beta_{qr}\) therefore measures the tendency of two transformations to occur together within a local context: positive values indicate stylistic reinforcement, whereas negative values indicate a trade-off between transformation types. Diagonal elements \(\beta_{qq}\) capture recurrence, reflecting the tendency of the same transformation family to appear repeatedly within a segment.
\end{itemize}

Together, the unary and pairwise components relate motif-level musical characteristics to transformation behavior while characterizing the structured co-occurrence of transformations within phrase-level contexts.

\subsection{Estimation via Pseudo-likelihood}
\label{subsec:pl_estimation}

Exact maximum likelihood estimation of the conditional model in \eqref{eq:crf} is computationally infeasible. Because motif instances within each segment form densely connected graphs, maximizing the likelihood requires evaluating the partition function, which involves summation over all possible configurations of the multilabel matrix \(Y\). This normalization term grows exponentially in the number of instances and transformation types, making exact computation intractable.

To obtain tractable parameter estimates, we use the pseudo-likelihood approximation \citep{Besag1975}, a standard approach for Conditional Random Fields with cyclic dependency structures \citep{Sutton2012}. Instead of maximizing the full likelihood \(P(Y \mid X)\), the pseudo-likelihood is defined as the sum of log-conditional probabilities for each transformation indicator given all remaining labels and the observed features:
\begin{equation}
\mathcal{L}_{\text{PL}}(\alpha,\beta)
=
\sum_{i=1}^{N} \sum_{q=1}^{Q}
\log P\!\left(Y_i^q \mid Y_{-(i,q)}, X\right),
\end{equation}
where \(Y_{-(i,q)}\) denotes all transformation indicators except the entry \(Y_i^q\).

Under the log-linear specification of the CRF, conditioning on all other labels cancels the global normalization term. The resulting conditional distribution takes a logistic form,
\begin{equation}
P(Y_i^q = 1 \mid Y_{-(i,q)}, X)
=
\sigma\!\left(
X_i^\top \alpha_{\cdot q}
+
\sum_{j=1}^{N} \mathcal{A}_{ij}
\sum_{r=1}^{Q} Y_j^r \beta_{qr}
\right),
\label{eq:pl_logistic}
\end{equation}
where \(\sigma(z) = (1+e^{-z})^{-1}\) and \(\alpha_{\cdot q} \in \mathbb{R}^p\) denotes the coefficient vector associated with transformation \(q\). Each transformation indicator therefore depends on motif-level features and on the graph-weighted configuration of transformations in nearby instances.

Parameter estimation is performed by maximizing the \emph{regularized pseudo-likelihood},
\begin{equation}
\mathcal{L}_{\text{PL,reg}}
=
\sum_{i=1}^{N} \sum_{q=1}^{Q}
\left[
Y_i^q z_{iq} - \log(1+e^{z_{iq}})
\right]
-
\lambda_\alpha \|\alpha\|_2^2
-
\lambda_\beta \|\beta\|_2^2,
\end{equation}
where
\begin{equation}
z_{iq}
=
X_i^\top \alpha_{\cdot q}
+
\sum_{j=1}^{N} \mathcal{A}_{ij}
\sum_{r=1}^{Q} Y_j^r \beta_{qr}.
\end{equation}
Small \(L_2\) penalties (\(\lambda_\alpha = \lambda_\beta = 10^{-3}\)) were included to stabilize optimization in the sparse multilabel setting by preventing overly large coefficients in the presence of correlated features. In practice, results were insensitive to the choice of small penalty values, and the estimates were numerically indistinguishable from those obtained without regularization.

Because log-linear graphical models admit multiple equivalent parameterizations, additional constraints were imposed to ensure a unique and interpretable solution. The interaction matrix \(\beta\) was constrained to be symmetric and centered so that its row and column sums equal zero, ensuring that interaction parameters reflect relative co-occurrence tendencies rather than arbitrary shifts. The feature coefficients in \(\alpha\) (excluding the bias term) were mean-centered to prevent confounding between feature effects and interaction terms. A constant bias column was included in the feature matrix so that baseline transformation frequencies are captured explicitly by intercept parameters, separating baseline effects (\(\alpha\)) from genuine transformation interactions (\(\beta\)).

The regularized objective was optimized using the L-BFGS quasi-Newton method \citep{Byrd1995}, which provides stable and efficient convergence for smooth problems of moderate dimensionality.

\subsection{Statistical Inference}
\label{subsec:statistical_inference}

Statistical inference was used to evaluate both the global contribution of model components and the uncertainty of individual parameters. Global model comparisons were conducted using permutation-based composite likelihood ratio (CLR) tests \citep{Good2005}, while parameter uncertainty was assessed using Wald confidence intervals based on the Godambe sandwich variance for composite likelihood estimators \citep{Varin2011}.

\subsubsection{Composite Likelihood Ratio Tests}

Model comparisons were designed to evaluate three distinct sources of musical structure represented in the CRF:

\begin{itemize}
\item \textbf{Baseline vs.\ unary model.}  
Does incorporating motif-level musical features improve the explanation of transformation behavior beyond baseline transformation frequencies?

\item \textbf{Unary vs.\ full model.}  
Do contextual dependencies between nearby motifs (pairwise effects) improve model fit beyond what can be explained by motif-level features alone?

\item \textbf{Pairwise vs.\ full model.}  
Do motif-level features provide additional explanatory power once baseline frequencies and contextual dependencies are already taken into account?
\end{itemize}

For each comparison, the test statistic was defined as the difference between the maximized pseudo-log-likelihood values of the alternative and nested models,
\begin{equation}
\mathrm{CLR}
=
\ell_{\mathrm{PL}}(\hat{\theta}_{\text{alt}})
-
\ell_{\mathrm{PL}}(\hat{\theta}_{\text{null}}),
\end{equation}
where \(\ell_{\mathrm{PL}}\) denotes the pseudo-log-likelihood and \(\hat{\theta}\) is the corresponding maximum pseudo-likelihood estimate.

Reference distributions for the CLR statistics were obtained using permutation tests, which provide an empirical finite-sample calibration under the null hypothesis \citep{Good2005}.

Permutations were performed separately within each stylistic period and constrained at the \emph{segment level}. This reflects the CRF design: segments form independent subgraphs and therefore constitute the appropriate independent observational units.

Let \(S = \mathcal{A}Y\) denote the matrix of graph-weighted neighboring transformation configurations. For each comparison, permutations were constructed to disrupt only the structure under test:

\begin{itemize}
\item \textbf{Baseline vs.\ unary.}  
Motif-level covariates in \(X\) were permuted within each segment, breaking the association between musical features and transformation labels while preserving segment structure.

\item \textbf{Unary vs.\ full.}  
The neighbor configuration matrix \(S\) was permuted within each segment, destroying contextual dependencies among motifs while leaving unary feature effects unchanged.

\item \textbf{Pairwise vs.\ full.}  
Motif-level covariates were permuted within each segment, testing whether unary musical features contribute beyond a model that already captures baseline transformation frequencies and contextual interactions.
\end{itemize}

For each test, the CLR statistic was recomputed over \(B\) permutations, and empirical p-values were calculated as
\begin{equation}
p^{\mathrm{perm}}
=
\frac{1 + \#\{\mathrm{CLR}^{(b)} \ge \mathrm{CLR}_{\mathrm{obs}}\}}{B + 1}.
\end{equation}

These tests therefore quantify whether motif-level characteristics and phrase-level context contribute meaningfully to the transformation patterns observed in the musical data.

\subsubsection{Wald Confidence Intervals}

Let \(\hat{\theta}\) denote the maximum pseudo-likelihood estimate obtained by stacking all unary coefficients \(\alpha\) and pairwise interaction parameters \(\beta\) into a single vector. Under standard regularity conditions, composite likelihood estimators are asymptotically normal \citep{Varin2011}. Their covariance is given by the Godambe sandwich matrix
\begin{equation}
\mathcal{G}(\hat{\theta})
=
\mathcal{H}^{-1}
\mathcal{J}
\mathcal{H}^{-1},
\end{equation}
where \(\mathcal{H}\) is the observed Hessian of the pseudo-log-likelihood and \(\mathcal{J}\) is the empirical covariance of the score vectors (i.e., gradients of the pseudo-log-likelihood evaluated at \(\hat{\theta}\)).

Standard errors were obtained from the diagonal elements,
\begin{equation}
\widehat{\mathrm{SE}}(\hat{\theta}_j)
=
\sqrt{\left[\mathcal{G}(\hat{\theta})\right]_{jj}},
\end{equation}
and confidence intervals were constructed using the asymptotic normal approximation.

In the present analysis, small regularization terms were included in the estimation procedure for numerical stability, but the resulting parameter estimates were numerically indistinguishable from those obtained without regularization. As a result, the reported standard errors and confidence intervals reflect the unregularized solution to a high degree of accuracy. More generally, when regularization has a non-negligible effect on the estimates, it may introduce bias, and care is required when interpreting inferential quantities; in such cases, it may be preferable to base inference on the unregularized objective or to consider bias-correction strategies.

Because unary and pairwise parameters are estimated jointly, the Godambe matrix captures their statistical dependence. For interpretation and presentation, inference was summarized separately for unary effects (\(\alpha\)) and pairwise interaction parameters (\(\beta\)) using the corresponding sub-blocks of \(\mathcal{G}(\hat{\theta})\). Within each parameter family, multiple comparisons were controlled using the Benjamini-Hochberg false discovery rate procedure \citep{Benjamini1995}.

\subsection{Period-Based Models and Effective Sample Size}
\label{subsec:period_based_models}

Statistical inference in the present framework depends on the amount of independent information available at the level of local musical context. Dependencies are defined only within segment graphs, and no edges are specified across segment boundaries. As a result, segments form independent subgraphs and constitute the appropriate independent observational units for inference.

The effective sample size (ESS) therefore reflects the number of informative segments rather than the total number of motif instances. For unary effects, the relevant ESS corresponds to the total number of segments in the period. For pairwise interactions, the ESS for a coefficient $\beta_{qr}$, which describes the contextual association between transformation families $q$ and $r$-is the number of segments in which at least one of these transformations occurs, since only those segments contribute information about their co-occurrence. In sparse multilabel settings, low activation rates can substantially reduce the effective information available for estimating specific parameters.

Methodological guidelines for cluster-based inference suggest that reliable sandwich-based Wald and likelihood-type inference typically requires on the order of 50 independent clusters, with small-sample adjustments recommended when the effective sample size falls between approximately 30 and 50 \citep{cameron2015,fay2001}. The present period-based analyses exceed this threshold; however, the inferential framework allows for $t$-based reference distributions when the effective sample size is moderate and normal approximations when it is sufficiently large.

Individual sonatas contain too few segments to meet these criteria. To obtain stable inference while preserving musical interpretability, models were therefore estimated at the stylistic-period level using the standard chronological division of Beethoven’s output \citep{BeethovenGrove2001}. The early period includes Sonatas Nos.\ 1-11 and 19-20, and the middle period includes Nos.\ 12-18 and 21-27. Within each period, segments from all sonatas were pooled and the CRF was fitted jointly. Sonatas were concatenated for convenience, but the graphical structure was unchanged: segments remained independent subgraphs, and no dependencies were introduced across segment or sonata boundaries.

The late period (Sonatas Nos.\ 28-32) was not included in the comparative analysis. Under the transformation scheme used in this study, several transformation families occur only rarely in these works, resulting in sparse label configurations and limited information for estimating multiple coefficients. This sparsity leads to low effective sample sizes for a number of pairwise interactions and unstable parameter estimates. Additional subsampling analyses indicate that this sparsity is not solely a consequence of the smaller number of sonatas in the late period, but also reflects differences in the frequency and structural organization of transformation types. In particular, subsampling experiments with comparable sample sizes in the early and middle periods yielded stable and statistically informative estimates, supporting the interpretation that the late-period sparsity reflects a structural property of the data rather than a limitation of sample size alone. Restricting the analysis to the early and middle periods therefore enables a balanced comparison based on sufficiently informative data.

\section{Results}

Results are reported separately for the early and middle periods, with emphasis on differences in transformation prevalence, feature associations, and contextual dependencies. We begin with a descriptive overview of the datasets and the distribution of transformation families, followed by statistical comparisons of the contributions of musical features and local context. The final analyses examine the estimated feature effects and interaction structure among transformation types across periods.

\subsection{Corpus Overview and Transformation Prevalence}

The early-period dataset included 13 piano sonatas, comprising 1,953 motif instances distributed across 242 segments. These works correspond to Beethoven’s early compositional phase, which is generally characterized by adherence to Classical formal conventions and relatively stable motivic organization.

The middle-period dataset included 14 sonatas, containing 2,192 motif instances across 204 segments. This phase is often associated in the literature with Beethoven’s so-called “heroic” style, characterized by expanded formal scope and intensified developmental processes, sometimes realized through large-scale structures and at other times through highly concentrated motivic development within shorter spans \citep{BeethovenGrove2001}. Although the late period is often described as exhibiting greater structural experimentation, it was not included in the comparative analysis due to the sparsity of transformation labels under the present modeling framework (see Section~\ref{subsec:period_based_models}).

Table~\ref{tab:label_prevalence} reports the prevalence of each transformation family by period. Overall, transformation labels are sparse, with most families occurring in fewer than one quarter of motif instances. Note addition/removal is the most frequent transformation in both periods (about 39\%), followed by salient-leap-preserving and intervallic-preserving transformations.

Comparisons between periods show modest differences in prevalence. The middle period exhibits higher frequencies of intervallic preservation (0.212 vs.\ 0.193) and symmetry (0.068 vs.\ 0.051), along with a slight increase in harmony-preserving transformations. In contrast, identity (0.049 vs.\ 0.056) and rhythm-preserving (0.120 vs.\ 0.135) occurrences are less frequent. These patterns are consistent with reduced reliance on literal or proportionally scaled repetition and a greater incidence of structurally modified realizations in the middle-period works.

\begin{table}
\tbl{Prevalence of transformation families by stylistic period. Frequencies are relative to the total number of motif instances in each period (Early: $N=1953$; Middle: $N=2192$).}
{\begin{tabular}{lcc} \toprule
Transformation & Early (freq.) & Middle (freq.) \\ \midrule
Identity        & 0.056 & 0.049 \\
Contour         & 0.157 & 0.157 \\
Salient Leap    & 0.229 & 0.225 \\
Rhythm          & 0.135 & 0.120 \\
Note Addition/Removal & 0.388 & 0.387 \\
Harmony         & 0.110 & 0.115 \\
Intervallic     & 0.193 & 0.212 \\
Symmetry        & 0.051 & 0.068 \\
\bottomrule
\end{tabular}}
\label{tab:label_prevalence}
\end{table}

\subsection{Global Model Comparison}

To evaluate the relative contribution of motif-level musical features and segment-level contextual dependencies, we compared nested CRF specifications using permutation-based composite likelihood ratio (CLR) tests (Section~\ref{subsec:statistical_inference}). Empirical p-values were obtained from 1,000 segment-level permutations. Table~\ref{tab:clr_results} summarizes the observed CLR statistics and permutation p-values for each comparison, reported separately by period.

In both periods, each nested comparison yields a significant improvement in pseudo-likelihood. The unary model provides a better fit than the baseline model, indicating that motif-level features contribute explanatory power beyond baseline transformation frequencies. Adding pairwise terms further improves model fit, demonstrating that contextual dependencies among nearby motif instances provide additional information beyond motif-level features. The full model also outperforms the pairwise specification, indicating that unary feature effects remain informative even when contextual interactions are included.

Across periods, the gain from baseline to unary is of similar magnitude, and the gain from pairwise to full is also comparable. In contrast, the improvement from unary to full is larger in the middle period than in the early period. This indicates that the addition of pairwise terms yields a larger increase in pseudo-likelihood in the middle-period data, indicating a larger relative contribution of contextual dependencies within the present modeling framework.

\begin{table}
\tbl{Permutation-based composite likelihood ratio (CLR) tests for nested model comparisons by stylistic period. $p^{\mathrm{perm}}$ denotes empirical p-values obtained from segment-level permutations (1,000 resamples).}
{\begin{tabular}{lcccc} \toprule
Comparison & Early CLR & Early $p^{\mathrm{perm}}$ & Middle CLR & Middle $p^{\mathrm{perm}}$ \\ \midrule
Baseline $\rightarrow$ Unary   & 477.1  & 0.0010 & 405.0  & 0.0010 \\
Unary $\rightarrow$ Full       & 926.9  & 0.0030 & 1299.7 & 0.0010 \\
Pairwise $\rightarrow$ Full    & 268.5  & 0.0010 & 275.6  & 0.0010 \\
\bottomrule
\end{tabular}}
\label{tab:clr_results}
\end{table}

\subsection{Unary Feature Effects}
\label{subsec:unary_effects}

Unary effects describe how local musical properties are associated with the occurrence of individual transformation families within each stylistic period. Tables~\ref{tab:unary_early}-\ref{tab:unary_middle} report coefficients that remain significant after false-discovery-rate correction (Benjamini-Hochberg; $q<0.05$).

\subsubsection{Early Period}

In the early-period works, several transformation families exhibit systematic associations with structural properties of the local musical setting. A prominent pattern concerns tonal stability: higher rates of local key change are associated with lower probabilities of multiple transformations, including contour-preserving, salient-leap-preserving, rhythm-preserving, harmony-preserving, and symmetry-related outcomes. Within the early-period data, motivic transformation is therefore less likely to occur in passages characterized by frequent local tonal motion.

A second group of effects relates to registral and temporal organization. Greater pitch spread is associated with increased likelihood of salient-leap-preserving, rhythm-preserving, and note-edit transformations, while showing a negative association with symmetry. In addition, higher motif register is negatively associated with symmetry, indicating that symmetry-related transformations occur more often at relatively lower pitch levels within the local tonal context. Silence proportion and metrical stress further differentiate repetition from variation: identity is more likely under higher metrical stress and lower silence proportion, whereas rhythm-preserving transformations are associated with lower metrical stress and higher silence proportion. In addition, secondary-function harmony shows a positive association with symmetry, indicating that symmetry-related transformations tend to occur more frequently in harmonically marked contexts.

After multiple-comparison correction, no unary effects are retained for harmonic-complexity spread, expressive density, or accentuation variability in the early-period model.

\begin{table}
\tbl{Unary feature effects ($\alpha$) in the early period. The table reports coefficients with Benjamini-Hochberg false discovery rate adjusted values $q^{BH}<0.05$. Estimates are shown with standard errors in parentheses and 95\% Wald confidence intervals.}
{\begin{tabular}{lllll} \toprule
Transformation & Feature & Estimate (SE) & 95\% CI & $q^{BH}$ \\ \midrule
Identity & Intercept & -2.918 (0.250) & [-3.409, -2.427] & 0.000 \\
Identity & Metrical Stress Rate & 0.209 (0.048) & [0.115, 0.303] & 0.000 \\
Identity & Silence Proportion & -0.264 (0.027) & [-0.318, -0.211] & 0.000 \\
Contour & Intercept & -1.682 (0.189) & [-2.052, -1.312] & 0.000 \\
Contour & Key Change Count & -0.148 (0.033) & [-0.213, -0.084] & 0.000 \\
Salient Leap & Intercept & -1.607 (0.242) & [-2.081, -1.134] & 0.000 \\
Salient Leap & Key Change Count & -0.191 (0.054) & [-0.296, -0.086] & 0.002 \\
Salient Leap & Pitch Spread & 0.197 (0.065) & [0.069, 0.325] & 0.011 \\
Rhythm & Intercept & -2.170 (0.186) & [-2.534, -1.806] & 0.000 \\
Rhythm & Key Change Count & -0.238 (0.045) & [-0.326, -0.149] & 0.000 \\
Rhythm & Metrical Stress Rate & -0.388 (0.092) & [-0.568, -0.208] & 0.000 \\
Rhythm & Pitch Spread & 0.251 (0.061) & [0.131, 0.371] & 0.000 \\
Rhythm & Silence Proportion & 0.109 (0.034) & [0.043, 0.176] & 0.006 \\
Note Addition/Removal & Intercept & -0.786 (0.290) & [-1.353, -0.218] & 0.028 \\
Note Addition/Removal & Pitch Spread & 0.271 (0.101) & [0.074, 0.468] & 0.028 \\
Harmony & Intercept & -2.175 (0.214) & [-2.595, -1.756] & 0.000 \\
Harmony & Key Change Count & -0.284 (0.108) & [-0.496, -0.073] & 0.033 \\
Intervallic & Intercept & -1.601 (0.164) & [-1.923, -1.279] & 0.000 \\
Symmetry & Intercept & -3.399 (0.212) & [-3.814, -2.984] & 0.000 \\
Symmetry & Key Change Count & -0.308 (0.017) & [-0.341, -0.274] & 0.000 \\
Symmetry & Motif Pitch Register & -0.381 (0.059) & [-0.496, -0.265] & 0.000 \\
Symmetry & Pitch Spread & -0.140 (0.036) & [-0.210, -0.071] & 0.000 \\
Symmetry & Secondary Chord Proportion & 0.346 (0.091) & [0.168, 0.523] & 0.001 \\
\bottomrule
\end{tabular}}
\label{tab:unary_early}
\end{table}

\subsubsection{Middle Period}

In the middle-period works, the retained unary effects emphasize associations with temporal stability, metric organization, and local surface conditions. In contrast to the early period, no effects are retained for key-change activity, suggesting that transformation probabilities do not show systematic associations with local tonal-change rates under the present feature set.

A first group of effects concerns temporal and articulatory stability. Greater inter-onset-interval variability is associated with lower probabilities of identity and intervallic-preserving transformations, while higher accentuation variability is associated with reduced likelihood of both identity and rhythm-preserving outcomes. Together, these patterns indicate that increased temporal or articulatory variability is associated with a reduced prevalence of structurally stable or literal motivic realizations.

Metric structure continues to play a stabilizing role. Higher metrical stress is positively associated with identity and with contour-preserving transformations, suggesting that strong metric anchoring is associated with both literal repetition and structurally stable contour realizations.

Several effects relate to the contextual role of silence. Greater silence proportion is associated with increased likelihood of identity and salient-leap-preserving transformations, but with a reduced likelihood of contour-preserving outcomes. These patterns suggest that local gaps are associated with the framing of both repetition and selective forms of variation rather than functioning primarily as a disruption.

Registral span also shows selective associations. Greater pitch spread is positively associated with contour-preserving and note-edit transformations, linking wider registral contexts with wider melodic spans associated with contour-preserving and note-edit transformations.

Finally, harmonic and expressive surface conditions show additional associations. Secondary-function harmony is negatively associated with identity and rhythm-preserving transformations, indicating lower prevalence of literal repetition and rhythmic stability in harmonically colored contexts. In contrast, expressive density is positively associated with rhythm-preserving transformations, suggesting increased rhythmic shaping in locally dense expressive environments.

After multiple-comparison correction, no unary effects are retained for harmonic-complexity spread, key-change activity, or motif pitch register in the middle-period model.

\begin{table}
\tbl{Unary feature effects ($\alpha$) in the middle period. The table reports coefficients with Benjamini-Hochberg false discovery rate adjusted values $q^{BH}<0.05$. Estimates are shown with standard errors in parentheses and 95\% Wald confidence intervals.}
{\begin{tabular}{lllll} \toprule
Transformation & Feature & Estimate (SE) & 95\% CI & $q^{BH}$ \\ \midrule
Identity & Intercept & -3.024 (0.086) & [-3.192, -2.856] & 0.000 \\
Identity & Accentuation SD & -0.130 (0.020) & [-0.169, -0.092] & 0.000 \\
Identity & Metrical Stress Rate & 0.252 (0.049) & [0.155, 0.348] & 0.000 \\
Identity & SD IOI & -0.286 (0.041) & [-0.365, -0.206] & 0.000 \\
Identity & Secondary Chord Proportion & -0.347 (0.093) & [-0.528, -0.165] & 0.001 \\
Identity & Silence Proportion & 0.091 (0.025) & [0.041, 0.141] & 0.002 \\
Contour & Intercept & -1.909 (0.173) & [-2.247, -1.570] & 0.000 \\
Contour & Metrical Stress Rate & 0.226 (0.077) & [0.075, 0.377] & 0.015 \\
Contour & Pitch Spread & 0.206 (0.077) & [0.055, 0.356] & 0.030 \\
Contour & Silence Proportion & -0.157 (0.021) & [-0.198, -0.115] & 0.000 \\
Salient Leap & Intercept & -1.623 (0.161) & [-1.939, -1.307] & 0.000 \\
Salient Leap & Silence Proportion & 0.139 (0.046) & [0.050, 0.229] & 0.011 \\
Rhythm & Intercept & -2.204 (0.094) & [-2.388, -2.020] & 0.000 \\
Rhythm & Accentuation SD & -0.139 (0.040) & [-0.217, -0.061] & 0.002 \\
Rhythm & Expressive Density & 0.201 (0.049) & [0.106, 0.297] & 0.000 \\
Rhythm & Secondary Chord Proportion & -0.213 (0.069) & [-0.348, -0.079] & 0.009 \\
Note Addition/Removal & Intercept & -0.872 (0.267) & [-1.395, -0.349] & 0.006 \\
Note Addition/Removal & Pitch Spread & 0.261 (0.091) & [0.083, 0.438] & 0.017 \\
Harmony & Intercept & -2.339 (0.160) & [-2.652, -2.026] & 0.000 \\
Intervallic & Intercept & -1.583 (0.162) & [-1.901, -1.266] & 0.000 \\
Intervallic & SD IOI & -0.161 (0.042) & [-0.244, -0.079] & 0.001 \\
Symmetry & Intercept & -2.800 (0.167) & [-3.128, -2.472] & 0.000 \\
\bottomrule
\end{tabular}}
\label{tab:unary_middle}
\end{table}

\subsubsection{Cross-Period Summary}

The cross-period comparison is based on separate models estimated for each period, with differences interpreted through comparison of the corresponding coefficients and retained effects in Tables~\ref{tab:unary_early} and~\ref{tab:unary_middle}.

Across periods, unary effect magnitudes are generally moderate and align with the prevalence patterns reported earlier, indicating gradual feature-related shifts in transformation likelihood rather than sharply selective feature signatures. The negative intercept values further reflect the overall sparsity of transformation events, with feature effects acting primarily as local modulations around low baseline probabilities. 

The principal contrast concerns the role of tonal stability: local key-change activity is associated with reduced probabilities of several transformation families in the early-period model but shows no retained effects in the middle-period model. Conversely, effects related to temporal variability and surface-level expressive conditions are more prominent in the middle-period results. Silence-related effects also shift in character, being associated primarily with reduced identity and increased rhythmic modification in the early period, but with both repetition and selective variation in the middle period.

These differences motivate the subsequent examination of pairwise dependency structure, which captures how transformation types co-occur within local contexts beyond their individual associations with motif-level features.

\subsection{Pairwise Transformation Dependencies}
\label{subsec:pairwise_effects}

Pairwise effects describe segment-level co-occurrence tendencies between transformation families, estimated across segments after accounting for motif-level feature effects. Tables~\ref{tab:pairwise_early}-\ref{tab:pairwise_middle} report significant pairwise coefficients ($\beta$ parameters). Positive coefficients indicate that two transformation types are more likely to occur together within a segment, conditional on the unary structure, whereas negative coefficients indicate reduced co-occurrence.

\subsubsection{Early Period}

The early-period pairwise effect structure is characterized by strong positive self-association for all transformation families. Each transformation type shows a tendency to cluster with itself within segments, indicating that segments often exhibit a dominant local transformation regime rather than a mixed configuration of transformation types.

Cross-family effects that survive multiple-comparison correction are all negative. Identity shows reduced co-occurrence with several structurally modifying transformations, most strongly with symmetry and harmony-preserving transformations, and more moderately with note addition/removal. These patterns suggest that segments characterized by literal repetition tend to occur in contexts that are distinct from those involving stronger structural reorganization.

Several additional negative associations reflect differentiation among types of structural modification. Contour-preserving transformations show reduced co-occurrence with intervallic-preserving, rhythm-preserving, and note-edit transformations, with the strongest contrast observed for the contour--intervallic pair. Rhythm-preserving transformations also show negative associations with intervallic-preserving and note-edit outcomes. Harmony-preserving transformations exhibit reduced co-occurrence with intervallic-preserving and symmetry-related transformations.

Overall, the early-period pairwise effect pattern reflects a structured pattern of negative associations among transformation types without indicating consistent positive co-occurrence across families. Segments tend to be internally homogeneous with respect to transformation family, while consistent co-occurrence of different structural principles is not observed at a statistically significant level.

\begin{table}
\tbl{Pairwise effects ($\beta$) in the early period. The table reports coefficients with Benjamini-Hochberg false discovery rate adjusted values $q^{BH}<0.05$. Estimates are shown with standard errors in parentheses and 95\% Wald confidence intervals.}
{\begin{tabular}{llll} \toprule
Pair & Estimate (SE) & 95\% CI & $q^{BH}$ \\ \midrule
Identity--Identity & 2.385 (0.789) & [0.838, 3.932] & 0.005 \\
Identity--Note Addition/Removal & -0.340 (0.052) & [-0.441, -0.239] & 0.000 \\
Identity--Harmony & -0.541 (0.137) & [-0.810, -0.273] & 0.000 \\
Identity--Symmetry & -1.473 (0.119) & [-1.707, -1.239] & 0.000 \\
Contour--Contour & 1.608 (0.305) & [1.011, 2.206] & 0.000 \\
Contour--Rhythm & -0.489 (0.128) & [-0.740, -0.239] & 0.000 \\
Contour--Note Addition/Removal & -0.592 (0.078) & [-0.745, -0.438] & 0.000 \\
Contour--Intervallic & -1.184 (0.106) & [-1.392, -0.976] & 0.000 \\
Salient Leap--Salient Leap & 1.827 (0.332) & [1.176, 2.478] & 0.000 \\
Salient Leap--Rhythm & -0.622 (0.178) & [-0.971, -0.273] & 0.001 \\
Rhythm--Rhythm & 2.658 (0.431) & [1.812, 3.503] & 0.000 \\
Rhythm--Note Addition/Removal & -0.319 (0.081) & [-0.477, -0.161] & 0.000 \\
Rhythm--Intervallic & -0.717 (0.087) & [-0.887, -0.546] & 0.000 \\
Note Addition/Removal--Note Addition/Removal & 1.457 (0.266) & [0.936, 1.979] & 0.000 \\
Note Addition/Removal--Intervallic & -0.546 (0.167) & [-0.873, -0.220] & 0.002 \\
Harmony--Harmony & 1.679 (0.416) & [0.863, 2.495] & 0.000 \\
Harmony--Intervallic & -0.473 (0.095) & [-0.659, -0.287] & 0.000 \\
Harmony--Symmetry & -0.674 (0.278) & [-1.218, -0.130] & 0.027 \\
Intervallic--Intervallic & 2.245 (0.405) & [1.451, 3.039] & 0.000 \\
Symmetry--Symmetry & 2.142 (0.605) & [0.957, 3.327] & 0.001 \\
\bottomrule
\end{tabular}}
\label{tab:pairwise_early}
\end{table}

\subsubsection{Middle Period}

The middle-period pairwise effect structure is also characterized by strong positive self-association across all transformation families. Each transformation type shows a tendency to cluster with itself within segments, indicating concentration of relatively homogeneous transformation regimes at the segment level.

Several positive cross-family effects are retained. Identity shows positive associations with contour-preserving transformations and with symmetry, indicating that literal repetition co-occurs with contour-preserving and symmetry-based transformations within segments.

At the same time, identity exhibits negative associations with all other transformation families. This pattern indicates that identity is selectively integrated with certain preservation-based relations while remaining separated from broader forms of structural modification.

Strong negative effects are also observed among several non-identity transformation pairs. Symmetry shows substantial negative co-occurrence with contour-preserving and harmony-preserving transformations. Additional contrasts appear between contour and rhythm, contour and intervallic preservation, and between salient-leap-preserving and intervallic-preserving or note-edit transformations. These patterns indicate continued differentiation among transformation principles within segments.

Taken together, the middle-period pairwise effect pattern combines strong within-family clustering with selective coexistence among certain preservation-based transformations, alongside persistent exclusions among structurally incompatible transformation types.

\begin{table}
\tbl{Pairwise effects ($\beta$) in the middle period. The table reports coefficients with Benjamini-Hochberg false discovery rate adjusted values $q^{BH}<0.05$. Estimates are shown with standard errors in parentheses and 95\% Wald confidence intervals.}
{\begin{tabular}{llll} \toprule
Pair & Estimate (SE) & 95\% CI & $q^{BH}$ \\ \midrule
Identity--Identity & 2.907 (0.734) & [1.469, 4.346] & 0.000 \\
Identity--Contour & 0.490 (0.078) & [0.337, 0.644] & 0.000 \\
Identity--Salient Leap & -0.789 (0.044) & [-0.875, -0.703] & 0.000 \\
Identity--Rhythm & -1.055 (0.061) & [-1.175, -0.936] & 0.000 \\
Identity--Note Addition/Removal & -0.436 (0.051) & [-0.535, -0.336] & 0.000 \\
Identity--Harmony & -1.085 (0.062) & [-1.208, -0.963] & 0.000 \\
Identity--Intervallic & -0.334 (0.071) & [-0.473, -0.194] & 0.000 \\
Identity--Symmetry & 0.301 (0.118) & [0.070, 0.532] & 0.017 \\
Contour--Contour & 2.145 (0.375) & [1.410, 2.881] & 0.000 \\
Contour--Rhythm & -0.880 (0.144) & [-1.162, -0.599] & 0.000 \\
Contour--Note Addition/Removal & -0.416 (0.111) & [-0.634, -0.198] & 0.000 \\
Contour--Intervallic & -0.372 (0.103) & [-0.574, -0.170] & 0.001 \\
Contour--Symmetry & -1.343 (0.171) & [-1.677, -1.009] & 0.000 \\
Salient Leap--Salient Leap & 2.471 (0.337) & [1.810, 3.132] & 0.000 \\
Salient Leap--Note Addition/Removal & -0.427 (0.102) & [-0.626, -0.227] & 0.000 \\
Salient Leap--Intervallic & -0.606 (0.203) & [-1.003, -0.208] & 0.005 \\
Rhythm--Rhythm & 1.982 (0.419) & [1.160, 2.803] & 0.000 \\
Rhythm--Intervallic & -0.268 (0.106) & [-0.475, -0.061] & 0.018 \\
Note Addition/Removal--Note Addition/Removal & 1.547 (0.395) & [0.773, 2.322] & 0.000 \\
Harmony--Harmony & 2.428 (0.462) & [1.522, 3.334] & 0.000 \\
Harmony--Symmetry & -1.445 (0.148) & [-1.734, -1.155] & 0.000 \\
Intervallic--Intervallic & 2.140 (0.444) & [1.269, 3.010] & 0.000 \\
Symmetry--Symmetry & 2.953 (0.689) & [1.602, 4.303] & 0.000 \\
\bottomrule
\end{tabular}}
\label{tab:pairwise_middle}
\end{table}

\subsubsection{Cross-Period Summary}

The cross-period comparison is based on separate models estimated for each period, with differences interpreted through comparison of the corresponding coefficients in Tables~\ref{tab:pairwise_early} and~\ref{tab:pairwise_middle}.

Across periods, the dominant feature of the pairwise effect structure is strong positive self-association for all transformation families, indicating that segments tend to concentrate a single transformation regime rather than exhibiting uniform mixing of multiple transformation types. Given the overall sparsity of transformation events, these effects are consistent with segment-level clustering beyond baseline prevalence differences.

The principal contrast concerns cross-family structure. In the early period, all retained cross-family effects are negative, reflecting a pattern in which cross-family co-occurrence is consistently negative across retained effects. In the middle period, this pattern changes: positive associations emerge between identity and certain preservation-based transformations, and the overall structure shows a broader range of both positive and negative cross-family relationships.

At the same time, several strong negative effects persist in the middle period, particularly involving symmetry and harmony-related transformations, indicating that structural differentiation remains an important organizing factor.

Beyond these observed negative effects, cross-period differences are reflected not only in the presence or absence of associations, but also in changes in their direction, strength, and selectivity across transformation pairs.

These results suggest a reorganization of cross-family interaction structure between periods. While the early period is characterized by consistently negative cross-family associations, the middle period exhibits a more heterogeneous pattern in which selective positive co-occurrence emerges alongside persistent strong exclusions. This indicates increased differentiation in how transformation types relate within local musical contexts, without eliminating the underlying tendency toward within-family clustering.

\section{Discussion}

This study examined how motifs are transformed within their musical context in Beethoven’s piano sonatas. Using a multilabel Conditional Random Field estimated for inferential purposes, we analyzed differences in transformation prevalence, retained feature associations, and the structure of contextual dependencies among transformation families across the early and middle stylistic periods. These findings provide a distributional characterization of transformation tendencies within local musical context rather than evidence for deterministic compositional rules.

The results indicate that transformation probabilities vary systematically with local musical properties and with the surrounding contextual configuration. Unary effects are generally moderate and act as adjustments around low baseline frequencies, pointing to gradual feature-related differences between periods rather than sharply selective signatures. At the same time, the pairwise structure indicates strong within-family co-occurrence across both periods, while cross-family relationships are more selective and often negative. In the early-period works, the pairwise structure is characterized by consistently negative cross-family associations, with no statistically significant positive co-occurrence between transformation families. In the middle-period works, this pattern is reorganized: selective positive associations emerge alongside strong negative relationships, reflecting changes in the direction, strength, and selectivity of cross-family interactions. This reorganization is also reflected in shifts in the magnitude of existing associations, as well as in the emergence of new positive and negative relationships across transformation pairs. Together, these observations suggest that differences between periods are reflected in how transformation types relate within local musical contexts, rather than in the overall presence or absence of segment-level organization, pointing to changes in the organization of motivic variation within local phrases.

The modeling approach provides a statistical setting for the analysis of annotated structural data organized into approximately independent contextual units. The CRF combines multilabel outcomes with an explicit representation of local dependence and pseudo-likelihood estimation, allowing both feature associations and contextual co-occurrence patterns to be evaluated with uncertainty quantification and multiple-comparison control. In this formulation, the model serves an analytical rather than predictive role, supporting the assessment of whether relationships suggested by exploratory analysis remain detectable once sparsity and local dependence are taken into account.

Although the empirical application focused on motivic transformations, the formulation is not specific to this domain. The transformation families, feature construction, and segment-based graph represent one operational design that enables inference. Alternative label definitions, including expert or perceptually motivated annotations, as well as alternative feature sets or local-context representations, can be incorporated within the same framework. More generally, any setting in which events are described by multilabel structural properties and organized into approximately independent local units may be examined in this way.

The analysis was restricted to the first movements of Beethoven’s piano sonatas and to the early and middle stylistic periods; the late period was excluded because several transformation families occurred too infrequently for stable estimation. Expanding the corpus would increase the number of informative contextual units and improve the stability of estimates for rare transformations and interactions. Such extensions would also allow comparisons across composers or repertoires. Within these limits, the present formulation offers a probabilistic approach to the study of structured annotations in symbolic corpora while accounting for local contextual dependence.

\section*{Disclosure statement}
The authors report no conflict of interest.

\bibliographystyle{apacite}
\bibliography{references}

\end{document}